\documentclass{article}
\usepackage[cp1251]{inputenc}
\usepackage[T2A]{fontenc}
\usepackage{amsmath}
\usepackage{indentfirst}
\usepackage{amssymb}
\usepackage{literat}
\usepackage{geometry}
\usepackage{graphicx}
\usepackage{cite}

\newcommand{\ve}{\varepsilon}
\newcommand{\MF}{\mathcal{F}}

\title{Properties of pattern formation and selection processes in nonequilibrium systems with external fluctuations}
\author{D.Kharchenko\footnote{dikh@ipfcentr.sumy.ua},
		V.Kharchenko,
		I.Lysenko\\
		\textit{Institute of Applied Physics, Nat.Acad. of Sci. of Ukraine,}\\ \textit{58
		  Petropavlovskaya St., 40030, Sumy, Ukraine}}

\begin{document}
\maketitle

\section*{Abstract}We extend the phase field crystal method for nonequilibrium patterning
to stochastic systems with external source where transient dynamics is
essential. It was shown that at short time scales the system manifests pattern
selection processes. These processes are studied by means of the structure
function dynamics analysis. Nonequilibrium pattern-forming transitions are
analyzed by means of numerical simulations.

\textbf{Key words:} noise, spatial pattern, nonequilibrium transition,
structure function.\\
 \textbf{PACS} 05.40.-a, 05.10.Gg, 64.60.Cn, 64.60.My

\section{Introduction}

A remarkable property of macroscopic systems is their ability to generate forms
or patterns. Numerous efforts have been devoted to study main principles of
pattern formation in last three decades. A considerable progress has been made
through the study of problems related to fluid flow, solidification processes,
formation of antiphase domain walls and grain boundaries, etc. The well known
examples of pattern formation are: convective rolls in Rayleigh-B\'enard cells
\cite{RBC1,RBC2,RBC3}, a Turing instability with spatio-temporal dynamics in
chemical systems \cite{Showalter}, formation of patterns on gelation surfaces
\cite{Katsuragi}, noise induced patterns in excitable systems
\cite{Sagues,Lindner} and formation of a semiconductor nanostructure
\cite{Stegemann}, noise induced and sustained patterns in reaction-diffusion
systems \cite{MHW2005,PhysD2009,MetPhys2009,PhysA_V2009,Mangioni}. Also pattern
formation processes can be induced by an external influence for example
irradiation \cite{Bellon2001,Bellon2004,LZWBE2006,KYH2009}.

A considerable study of microstructure transformation and patterning has been
given using phase field theory. In such an approach a local atomic mass density
field $\rho=\rho(\mathbf{r},t)$ is introduced to describe the phases present.
The standard phase field theory considers the field dynamics at diffusive time
and length scales. Recently, a new direction in phase field theory known as
phase field crystals method has been proposed \cite{EKHG_PRL2002}. This
approach allows one to simulate materials on microscopic scales (defects and
grain boundary formation) and effectively consider dynamics on diffusive scales
(solute transport). An advantage of this method is its ability to model a field
in the solid phase that exhibits the periodic nature, in the framework of this
approach different crystal-structure-related properties naturally arise in this
model \cite{Elder2009}. Moreover, this approach incorporates elastic and
plastic behaviour of periodic systems \cite{Elder2006}. Originally, the model
was postulated in the framework of Ginzburg-Landau theory of phase transitions
where a gradient expansion of the free energy is not limited by the first
non-vanishing term. The free energy functional incorporates spatial derivatives
of the field of higher order in a form of the Swift-Hohenberg operator
\cite{SH77} for spatial coupling that naturally describes periodic patterns due
to minimization of the free energy.

In the framework of the standard phase field crystals approach usually a
diffusive atomic dynamics is considered. Such slow dynamics can be observed in
systems near its equilibrium states with an instant response for the change in
the order parameter. In non-equilibrium systems it realizes if the time scale
for the observed phenomenon is larger than the transient period (systems with
dissipative dynamics). An approximation of the slow dynamics admits that a
propagation speed of disturbances is infinitely large. Generally, if one
considers systems out of equilibrium, then the transient dynamics should be
taken into account to satisfy the criterion of a finite propagation speed
according to physical microscopic processes. It follows, that the (fast)
dynamics of such systems can be studied on the time scales of the order or
smaller than the relaxation time. The modified phase field crystal method
introduced in Ref.\cite{SHP2006} includes both diffusive dynamics and elastic
interactions, where the separation of time scales exists between diffusive and
relaxation processes in solid.

The study of the pattern formation within this approach was made in a
deterministic (noiseless) limit. It was assumed that fluctuation sources
(noise) can not principally change the system dynamics and its stationary
states. Overdamped stochastic systems with nonconserved dynamics and the
Swift-Hohenberg spatial coupling were well studied in last two decades (see,
for example, \cite{EVG92,GHS93,GS94,GS96,PVBR96,ZS98,WBL06}). It was shown that
external fluctuations (additive or multiplicative) can induce pattern formation
in such systems. Unfortunately, the patterning processes in stochastic systems
with conserved dynamics where transient dynamics is essential were not
discussed yet.

In this paper we extend the mechanism of nonequilibrium patterning by
consideration of stochastic systems with fast dynamics where the density filed
is a conserved quantity. Starting from the balance equation we consider the
system with ordinary thermally sustained flux related to fast dynamics and
external influence leading to additional (athermal) atomic mixing. We assume
that every flux has its own fluctuations. Considering transient dynamics we
shall show that at early stages the system manifests pattern selection
processes. We discuss nonequilibrium external noise-induced pattern-forming
transitions in such systems.

The paper is organized as follows. We introduce the model in Section
\ref{sec1}. The short time instability analysis and a possibility of pattern
selection processes are discussed in Section \ref{sec2}. Analytical results we
compare with computer simulations in Section \ref{sec3} where nonequilibrium
pattern-forming transitions are studied. Finally, we summarize our main
conclusions in Section \ref{sec5}.

\section{Model}\label{sec1}

Let us consider a class of extended systems described by a scalar conserved
field $x(\mathbf{r},t)$ obeying the mass conservation law $\int{\rm d}
\mathbf{r}x(\mathbf{r},t)=const$. The field $x$ relates to the mass density
field $\rho(\mathbf{r},t)$ as $x=(\rho-\rho_0)/\rho_0$, where $\rho_0$ is a
uniform reference density. A generic evolution model of the field $x$ is given
by the continuity equation
\begin{equation}\label{eq1}
\partial_t x=-\nabla\cdot \mathbf{J}_{tot}.
\end{equation}
Here $\mathbf{J}_{tot}$ is a total flux which generally depends on the time
$t$. Let us assume that $\mathbf{J}_{tot}$ consists of a thermally sustained
diffusion flux $\mathbf{J}_{D}$ and an atomic mixing flux $\mathbf{J}_{e}$
induced by external influence (leading to a structural disorder, turbulence
effects, etc.), so that $\mathbf{J}_{tot}= \mathbf{J}_{D}+ \mathbf{J}_{e}$. The
flux $\mathbf{J}_{e}$ emerges as a result of atomic mixing caused for example
by interactions of irradiated high-energy particles with atoms of the system
(the ballistic flux).

Examining the system in real conditions we assume that every constituent of the
total flux has both regular and stochastic components. The thermally sustained
diffusion flux $\mathbf{J}_{D}$ is assumed to be described by a relaxational
Maxwell-Cattaneo equation \cite{JCL} generalized by a stochastic contribution
$\xi$, representing flux fluctuations. Therefore, the time evolution of the
diffusion flux is governed by the Langevin equation of the form
\cite{JCL,Galenko2009,PhysA2008}
\begin{equation}\label{eq2}
\tau_D\partial_t\mathbf{J}_{D}=- \mathbf{J}_{D} -M\nabla \frac{\delta
\mathcal{F}}{\delta
 x}+\xi(\mathbf{r},t),
\end{equation}
were $\tau_D$ and $M=const$ are the relaxation time and the atomic mobility,
respectively; $\mathcal{F}$ is the free energy functional of the system;
$\xi(\mathbf{r},t)$ is a Gaussian noise representing thermal fluctuations with
$\langle\xi(\mathbf{r},t)\rangle=0$ obeying the fluctuation dissipation
relation $\langle\xi(\mathbf{r},t)\xi(\mathbf{r}',t)\rangle=2\sigma^2_0 M
\delta(\mathbf{r}-\mathbf{r}')\delta(t-t')$, $\sigma^2_0$ is the noise
intensity reduced to the bath temperature $T$. The relaxation term reflects the
memory effects; it is dominant at fast but finite speed of propagation
$v_D=\sqrt{M/\tau_D}$. In the case $\tau_D\to 0$ the diffusion flux takes the
``usual'' form $\mathbf{J}_{D}\simeq -M\nabla\delta \mathcal{F}/\delta
x+\xi(\mathbf{r},t)$ as a result of an instant response with the diffusive
speed $v_D\to\infty$. While the real diffusion processes occur at finite speed
$v_D$, next we consider the case $\tau_D\ne0$.

We study the model where the external flux $\mathbf{J}_{e}$ satisfies the Fick
law of the form $\mathbf{J}_{e}=-D^0_{e}\nabla x$, $D^0_{e}$ is the external
source induced effective diffusion coefficient. If one assumes that this
external influence has stochastic nature (for example irradiated high-energy
particles have Maxwell distribution with spatial correlations \cite{Ponomarev})
then one can put $D^0_{e}=D_{e}+\zeta(\mathbf{r},t)$, where $D_{e}$ is the
regular part of the ballistic flux $\mathbf{J}_{e}$, and $\zeta(\mathbf{r},t)$
is its stochastic component describing fluctuations that emerge in collision
processes in the system \cite{Yanovski}. Considering the general problem for
the external noise $\zeta(\mathbf{r},t)$ we adopt the following properties:
\begin{equation}\label{cz}
\begin{split}
\langle\zeta(\mathbf{r},t)\rangle=0,\quad
\langle\zeta(\mathbf{r},t)\zeta(\mathbf{r}',t)\rangle=2D_{e}\sigma^2
C(\mathbf{r}-\mathbf{r}')
 \delta(t-t');\\
C(\mathbf{r}-\mathbf{r}')=\frac{1}{(\sqrt{2\pi}r_c)^{d}}\exp\left(-\frac{(\mathbf{r}-\mathbf{r}')^2}{2r_c^2}\right)
\end{split}
\end{equation}
where $d$ is the spatial dimension; $\sigma_e^2$ is the noise intensity; the
coefficient $D_{e}$ standing in the correlator (\ref{cz}) means that the
stochastic part of the ballistic flux $\mathbf{J}_{e}$ emerges only if $D_e\ne
0$. Here $r_c$ is the correlation radius of the external fluctuations.
Physically it corresponds to overlapping length of disturbed domains of atomic
configurations emerging as a results of the external influence. In case $r_c\to
0$ we arrive at the white noise assumption with no overlapping. Considering a
simplest case we assume that $\sigma^2$ and $r_c$ are independent parameters.

In the following study we consider the free energy in the standard form of the
phase field crystal models \cite{EKHG_PRL2002}, i.e.,
\begin{equation}\label{Fe}
\mathcal{F}=\int{\rm d}\mathbf{r}\left(f(x)+\frac{1}{2}(x\mathcal{L}x)\right),
\end{equation}
where $f(x)$ is the free energy density, $\mathcal{L}$ is the spatial coupling
operator. Next we investigate a class of the systems which are able to produce
periodic states due to minimization of the free energy $\mathcal{F}$. As it was
shown in Ref.\cite{ElderGrant2004} in order to construct the free energy
functional for periodic systems one needs to take into account that in the
lowest-order gradient expansion the coefficient of $|\nabla x|^2$ must be
negative. The corresponding inifinite gradients in $x$ need to be compensated
by introduction the next-order terms, $|\nabla^2 x|^2$. The next criterion for
periodic pattern formation is a double well form of the free energy density
$f(x)$; we take it in the form
\begin{equation}\label{f}
 f(x)=\frac{\ve}{2} x^2+\frac{x^4}{4},
\end{equation}
where $\ve=\theta-1$ is the control parameter defined through the temperature
$T$, counted off a critical mean filed value $T_c$, $\theta=T/T_c$. The spatial
coupling operator $\mathcal{L}$ leading to formation of periodic patterns of
the field $x$ with a fixed wave-number has the form of the Swift-Hohenberg
spatial interaction, $\mathcal{L}=(q_0^2+\nabla^2)^2$, where $q_0=2\pi/a$ is
the wave-number giving the minimum of $\mathcal{F}$, $a$ is the equilibrium
lattice spacing. With $\varepsilon<0$ the homogeneous state $x$ is unstable to
the formation of a periodic structure for some values of the wave-vector
$\mathbf{q}_0$. Next, we put $q_0=1$ for convenience. As it was shown in
Ref.\cite{EKHG_PRL2002,ElderGrant2004} a mathematical construction for
$\mathcal{L}$ incorporates elastic effects \footnote{An equivalent and
alternative representation for the free energy functional (\ref{Fe}) is
$\mathcal{F}=\int{\rm d}\mathbf{r}\left[
f(x)+\frac{1}{2}([1+\nabla^2]x)^2\right]$.}. Indeed, rewriting it in the form
of the gradient expansion one has $\mathcal{F}=\int{\rm
d}\mathbf{r}\left(f(x)+\left[-\beta |\nabla x|^2+\gamma |\nabla^2
x|^2\right]\right)$, where $\beta$ and $\gamma$ are some phenomenological
constants. Next, if we put $x=A\sin(2\pi r/ a)$, where $\beta=1/\pi^2$,
$\gamma= 8/a_0^2$ and substitute it into $\mathcal{F}$, then we get
$\mathcal{F}/a\simeq a^{-1}\int{\rm d}r f(x) - A^2/a_0^2+(4
A^2/a_0^4)\epsilon^2$, where $\epsilon\equiv a-a_0$. This free energy is
minimized at $a=a_0$, where $a_0$ sets periodicity of the system. The term
$\propto \epsilon^2$ determines an elastic energy (the Hook law). Than, $\beta$
and $\gamma$ relate to elastic constants \cite{EKHG_PRL2002,EPBSG2007,BEG2008}.

As a results the complete set of the dynamical equations describing the system
under stochastic influence takes the form
\begin{equation}\label{XJ}
\begin{split}
 &\partial_t x=-\nabla \mathbf{J}_{D}+D_e\Delta x+\nabla(\zeta\nabla x),\\
 &\tau_D \partial_t \mathbf{J}_{D}=-\mathbf{J}_{D}-M\nabla\frac{\delta \mathcal{F}}{\delta
 x}+\xi.
\end{split}
\end{equation}
If one consider the limiting case of $D_e=0$ (i.e., $\mathbf{J}_e=0$), then
combining equation for both the field $x$ and the flux $\mathbf{J}_D$ one
arrives at the phase field crystals model: $\tau_D\partial^2_{tt}x+\partial_t
x=\nabla\cdot (M\nabla\frac{\delta\mathcal{F}}{\delta x}+\xi)$. It follows that
a hyperbolic transport here (the time derivative of the second order) emerges
when relaxation processes of the diffusion flux are possible
\cite{Galenko2009}. As it was stated in previous studies this model allows one
to capture a dynamics of the system at time and space scales related to
molecular dynamics simulations ($\sim 10^{-12}s$, $\sim 10^{-9}m$) and consider
dynamics at diffusion time and space scales. We generalize such the model by
taking into account external stochastic influence with an assumption of an
immediate response to the external disturbance at a distinct point. In such a
case the presented model effectively takes into account microstructure
transformation processes in an extended window for the time and space scales.

It is known that the system with a hyperbolic transport can manifest pattern
selection processes at early stages of the system evolution \cite{lzg}. Our
model has the same properties and should manifest pattern selection. Therefore,
in further study we aim to investigate pattern selection processes in periodic
systems described by the model (\ref{XJ}) under external influence. We shall
consider the external noise induced ordering processes in the model where
regular and random parts of the external flux have competing contributions into
the system dynamics.

\section{Analysis of pattern selection processes} \label{sec2}

Considering the stochastic system one should note that only statistical
measurable quantities of the stochastic filed are informative: the average
$\langle x(\mathbf{r},t)\rangle$ (a volume fraction of the system component)
and the structure function $S_\mathbf{k}(t)$ (a Fourier transform of the two
point correlation function $\langle \delta x(\mathbf{r},t)\delta
x(\mathbf{r}',t)\rangle$, $\delta x=x-\langle x\rangle$), where $\mathbf{k}$ is
the wave-vector. To discuss a behaviour of both $\langle
x(\mathbf{r},t)\rangle$ and $S_\mathbf{k}(t)$ we obtain dynamical equations for
these quantities and analyze their solutions.

\subsection{Dynamics of the average}
To obtain a dynamical equation for the quantity $\langle x\rangle$ we average
the system (\ref{XJ}) over fluctuations and arrive at
\begin{equation}
\begin{split}
 &\partial_ t\langle x\rangle=-\nabla \langle \mathbf{J}_D\rangle+D_e\Delta \langle x\rangle+\nabla\langle \zeta\nabla x\rangle,\\
 &\tau_D\partial_t\langle \mathbf{J}_D\rangle=-\langle \mathbf{J}_D\rangle-\nabla M\left\langle\frac{\delta \MF}{\delta
 x}\right\rangle.
\end{split}
\end{equation}
Noise correlators in the first equation can be decomposed using the Novikov
theorem \cite{Novikov}
\begin{equation}\label{ExN}
\langle\zeta\nabla
x\rangle=D_{e}\tilde\sigma^2\int_{-\infty}^{\infty}C(\mathbf{r-r'})\nabla\left<\frac{\delta
x(\mathbf{r},t)}{\delta \zeta(\mathbf{r}',t)}\right>{\rm d}\mathbf{r}'.
\end{equation}
The response function in r.h.s. of Eq.(\ref{ExN}) can be computed from the
formal solution of the Langevin equation (\ref{XJ}) for the field $x$:
\begin{equation}\label{eqD}
\frac{\delta x(\mathbf{r},t)}{\delta
\zeta(\mathbf{r}',t)}=\nabla\left(\delta(\mathbf{r-r'})\nabla
x(\mathbf{r},t)\right).
\end{equation}
Substituting Eq.(\ref{eqD}) into Eq.(\ref{ExN}), we get \cite{Yanovski,Garcia}
\begin{equation}
\begin{split}
 &\langle\zeta\nabla
x\rangle=D_{e}\sigma_e^2\left[\left.C(\mathbf{r-r'})\right|_{\mathbf{r=r'}}\nabla^3\langle
x\rangle+\right.\\ &\left.2\left(\left.\nabla
C(\mathbf{r-r'})\right|_{\mathbf{r=r'}}\right)\nabla^2\langle x\rangle+(\nabla
\langle x\rangle)\nabla^2\left.C(\mathbf{r-r'})\right|_{\mathbf{r=r'}}\right].
\end{split}
\end{equation}
It should be noted that $C(\mathbf{r-r'})$ takes the maximal value at
$\mathbf{r=r'}$ that gives
\begin{equation}
\left.\nabla C(\mathbf{r-r'})\right|_{\mathbf{r=r'}}=0;\quad
\nabla^2\left.C(\mathbf{r-r'})\right|_{\mathbf{r=r'}}<0.
\end{equation}
Hence, introducing notation $\varpi(\nabla^2)=\ve+\mathcal{L}^2+3x_0^2$ with
$M=1$ where $x_0$ represents the homogeneous state, we arrive at the system
\begin{equation}\label{mXJ}
\left\{
\begin{split}
 &{\partial_t}\langle x\rangle=-\nabla \langle \mathbf{J}_D\rangle+D_e\Delta \langle x\rangle
 +D_e\sigma^2(\nabla^2C(|\mathbf{r}|)|_{\mathbf{r}=\mathbf{0}}\Delta\langle x\rangle +D_e\sigma^2C(\mathbf{0})\nabla^4\langle x\rangle\\
 &\tau_D{\partial_t}\langle \mathbf{J}_D\rangle=-\langle
 \mathbf{J}_D\rangle-\nabla\varpi(\nabla^2)\langle x\rangle.
\end{split}
\right.
\end{equation}
In our further study we move to the Fourier space. To that end let us introduce
$\langle x_\mathbf{k}(t)\rangle=\int{\rm d}\mathbf{r} \langle
x(\mathbf{r},t)\rangle e^{i\mathbf{k}\mathbf{r}}$, $\langle
\mathbf{J}_\mathbf{k}(t)\rangle=\int{\rm d}\mathbf{r} \langle
\mathbf{J}(\mathbf{r},t)\rangle e^{i\mathbf{k}\mathbf{r}}$ and rewrite
Eq.(\ref{mXJ}) in the form
\begin{equation}\label{kXJ}
\left\{
\begin{split}
 &\frac{{\rm d} \langle x_\mathbf{k}\rangle}{{\rm d}t}=-i\mathbf{k} \langle \mathbf{J}_{D\mathbf{k}}\rangle-D_e|\mathbf{k}|^2
 \langle x_\mathbf{k}\rangle -
   D_e\sigma^2\nabla^2C(\mathbf{r})|_{\mathbf{r}=\mathbf{0}} k^2\langle x_\mathbf{k}\rangle +
   D_e\sigma^2C(\mathbf{0})k^4\langle x_\mathbf{k}\rangle\\
 &\tau_D\frac{{\rm d}\langle \mathbf{J}_{D\mathbf{k}}\rangle}{{\rm d}t}=-\langle
 \mathbf{J}_{D\mathbf{k}}\rangle-i\mathbf{k}\varpi(k^2)\langle x_\mathbf{k}\rangle.
\end{split}
\right.
\end{equation}
The system of ordinary differential equations (\ref{kXJ}) has an analytical
solution.

Stability analysis of the homogeneous state $x_0$ can be performed for the
average $\langle x_\mathbf{k}\rangle$ in the simplest way. Let us differentiate
the first equation from the system (\ref{kXJ}) over the time $t$. Hence,
expressing the flux $\mathbf{J}_{D\mathbf{k}}$ from the first equation and
using time derivative of the flux from the second one, we finally obtain
\begin{equation}\label{Xk2}
 \tau_D\frac{{\rm d}^2\langle x_\mathbf{k}\rangle }{{\rm
d}t^2}=-\left(1+\tau_D D_e k^2\Xi(k^2)\right)\frac{{\rm d}\langle
x_\mathbf{k}\rangle }{{\rm d}t}-k^2\left(D_e\Xi(k^2)+\varpi(k^2)\right)\langle
x_\mathbf{k}\rangle,
\end{equation}
where the notation $\Xi(k^2)\equiv 1+\sigma^2(\nabla^2 C(|r|)_{r=0}-C(0)k^2)$
is introduced. A solution of the derived equation can be found in the form
$\langle x_{|\mathbf{k}|}(t)\rangle=\langle x_{|\mathbf{k}|}(0)\rangle
\exp(\phi(k)t)$. Inserting this solution into Eq.(\ref{Xk2}) we get an
expression for the phase
\begin{equation}\label{Phi(k)}
\phi(k)_\pm=-\frac{1+\tau_D D_ek^2\Xi(k^2)}{2\tau_D}
 \pm\frac{1}{2\tau_D}\sqrt{(1+\tau_D
 D_ek^2\Xi(k^2))^2-4\tau_Dk^2\left(D_e\Xi(k^2)+\varpi_\nu(k^2)\right)}.
\end{equation}
It is seen that the phase can have real and imaginary parts, i.e., $\phi(k)=\Re
\phi(k)+i\Im\phi(k)$. It follows that an unstable modes are possible only if
$\Re\phi(k)_+>0$. It is known that in systems with the Swift-Hohenberg
interaction there is the interval $k_{c1}\le k\le k_{c2}$, where
$\Re\phi(k)_+>0$ with $k_{c1}, k_{c2}\ne0$. In other words the first unstable
mode will always have finite period given by the wave-number from this
interval. The quantity $\Re\phi(k)_+>0$ has a unique peak always; its position
sets the most unstable mode with the wave-number $k_m$. From the expression
(\ref{Phi(k)}) it follows that the imaginary part of the phase emerges if the
condition $(1+\tau_D
D_ek^2\Xi(k^2))^2<4\tau_Dk^2\left(D_e\Xi(k^2)+\varpi_\nu(k^2)\right)$ is
satisfied. Therefore, the evolution of the average $\langle
x_{|\mathbf{k}|}(t)\rangle$ can be characterized by decaying oscillations with
the frequency $\Im\phi(k)$ and the decrement $(1+\tau_D
D_ek^2\Xi(k^2))/2\tau_D>0$. The domain for decaying perturbations is defined by
the condition $1+\tau_D D_ek^2\Xi(k^2)>0$. A domain for the stable modes is
limited by the wave-number
\begin{equation}
k_d^2=\frac{1}{2\sigma^2C(0)}\left(1+\sigma^2 \nabla^2
C(|r|)_{r=0}+\sqrt{(1+\sigma^2 \nabla^2
C(|r|)_{r=0})^2+\frac{4\sigma^2C(0)}{\tau_DD_e}}\right).
\end{equation}
A domain where the oscillating behaviour of $\langle
x_{|\mathbf{k}|}(t)\rangle$ is realized can be defined by solutions
$k_0=k_0(\theta, D_e,\sigma^2, r_c)$ of the equation
\begin{equation}
1-2\tau_Dk^2(D_e\Xi(k^2)+2\varpi_\nu(k^2))+(\tau_DD_ek^2\Xi(k^2))^2<0.
\end{equation}

\subsection{Dynamics of the structure function}

Let us obtain the dynamical equation for the structure function $S_\mathbf{k}$.
To that end we rewrite the system (\ref{XJ}) in the Fourier space:
\begin{equation}\label{sXJk}
\begin{split}
 &\frac{{\rm d}x_\mathbf{k}}{{\rm d}t}=-i\mathbf{k} \mathbf{J}_{D\mathbf{k}}-k^2D_e x_\mathbf{k}-k^2\zeta_\mathbf{k}x_\mathbf{k}\\
 &\tau_D\frac{{\rm d}\mathbf{J}_{D\mathbf{k}}}{{\rm
 d}t}=-\mathbf{J}_{D\mathbf{k}}-i\mathbf{k}\varpi(k^2)x_\mathbf{k}+\xi_\mathbf{k}.
\end{split}
\end{equation}
Than the equation for the structure function takes the form
\begin{equation}\label{S_k}
\frac{{\rm d}S_\mathbf{k}}{{\rm d}t}=-i\mathbf{k}\langle
\mathbf{J}_{D\mathbf{k}}x_{-\mathbf{k}}\rangle+i\mathbf{k}\langle
\mathbf{J}_{D-\mathbf{k}}x_{\mathbf{k}}\rangle-2k^2D_eS_\mathbf{k}-k^2(\langle\zeta_\mathbf{k}x_\mathbf{k}x_{-\mathbf{k}}\rangle
 +\langle\zeta_{-\mathbf{k}}x_{-\mathbf{k}}x_{\mathbf{k}}\rangle).
\end{equation}
The corresponding correlators can be obtained from the equation
\begin{equation}\label{J_k}
\tau_D\frac{{\rm d}\langle \mathbf{J}_{D\mathbf{k}}x_{-\mathbf{k}}\rangle}{{\rm
d}t}=-\langle
\mathbf{J}_{D\mathbf{k}}x_{-\mathbf{k}}\rangle-i\mathbf{k}\varpi(k^2)S_\mathbf{k}+\langle\xi_\mathbf{k}x_{-\mathbf{k}}\rangle.
\end{equation}
The system of two differential equation of the first order (\ref{S_k},
\ref{J_k}) can be rewritten in the form of the one equation of the second order
\begin{equation}
\begin{split}
\tau_D\frac{{\rm d}^2S_\mathbf{k}}{{\rm d}t^2}=&-(1+2k^2\tau_D D_e)\frac{{\rm
d}S_\mathbf{k}}{{\rm d}t}-2k^2(D_e+\varpi(k^2))S_\mathbf{k}\\
 &-k^2(\langle\zeta_\mathbf{k}x_\mathbf{k}x_{-\mathbf{k}}\rangle
 +\langle\zeta_{-\mathbf{k}}x_{-\mathbf{k}}x_{\mathbf{k}}\rangle)-i\mathbf{k}(\langle\xi_\mathbf{k}x_{-\mathbf{k}}\rangle+
  \langle\xi_{-\mathbf{k}}x_{\mathbf{k}}\rangle)\\
 &-k^2\tau_D\frac{{\rm
d}}{{\rm d}t}(\langle\zeta_\mathbf{k}x_\mathbf{k}x_{-\mathbf{k}}\rangle
 +\langle\zeta_{-\mathbf{k}}x_{-\mathbf{k}}x_{\mathbf{k}}\rangle).
\end{split}
\end{equation}
Decomposing the noise correlators with the help of the Novikov theorem with
\begin{equation}
\begin{split}
&\langle\zeta_{-\mathbf{k}}x_{-\mathbf{k}}x_{\mathbf{k}}\rangle=\langle\zeta_\mathbf{k}x_\mathbf{k}x_{-\mathbf{k}}\rangle=D_e\sigma^2(\nabla^2C(|r|)_{r=0}-C(0)k^2)S_\mathbf{k},\\
&\langle\xi_{-\mathbf{k}}x_{\mathbf{k}}\rangle=\langle\xi_\mathbf{k}x_{-\mathbf{k}}\rangle=i\mathbf{k}\theta,
\end{split}
\end{equation}
we arrive at the dynamical equation in the form
\begin{equation}\label{Skt}
\begin{split}
\tau_D\frac{{\rm d}^2S_\mathbf{k}}{{\rm d}t^2}=&-(1+2k^2\tau_D
D_e\Xi(k^2))\frac{{\rm d}S_\mathbf{k}}{{\rm
d}t}-2k^2(D_e\Xi(k^2)+\varpi(k^2))S_\mathbf{k}+2\theta k^2\\
 &-\frac{2k^2D_e\sigma^2}{(2\pi)^d}\int{\rm d}\mathbf{k}'C(|\mathbf{k}-\mathbf{k}'|)S_{\mathbf{k}'}(t)-
 \frac{2k^2\tau_DD_e\sigma^2}{(2\pi)^d}\int{\rm d}\mathbf{k}'C(|\mathbf{k}-\mathbf{k}'|)\frac{{\rm d}S_{\mathbf{k}'}(t)}{{\rm
d}t},
\end{split}
\end{equation}
where $\sigma^2_0=\theta$. It is seen that the obtained equation (\ref{Skt})
admits a solution of the form $S\propto e^{\varphi(k)t}$, where the phase
$\varphi(k)$, in general, can have real and imaginary parts,
$\varphi=\Re\varphi(k)+\Im\varphi(k)$. In the spinodal decomposition theory
with the hyperbolic transport ($\tau_D\ne 0$) \cite{Galenko2009}, where the
spatial interaction is governed by the term $|\nabla x|^2$ the real part
 $\Re\varphi(k)_+$ is known as an amplification rate
$R(k)=-\Re\varphi(k)_+$, where $S\propto e^{-R(k)t}$; the imaginary part
$\Im\varphi(k)$ is responsible for pattern selection processes. Next, we
discuss behaviour of the phase $\phi(k)$ in order to study main properties of
the system dynamics, whereas the pattern selection processes we study
considering the structure function dynamics.

\subsection{The external noise influence on the pattern selection}

\begin{figure}
\centering
 a)\includegraphics[width=60mm]{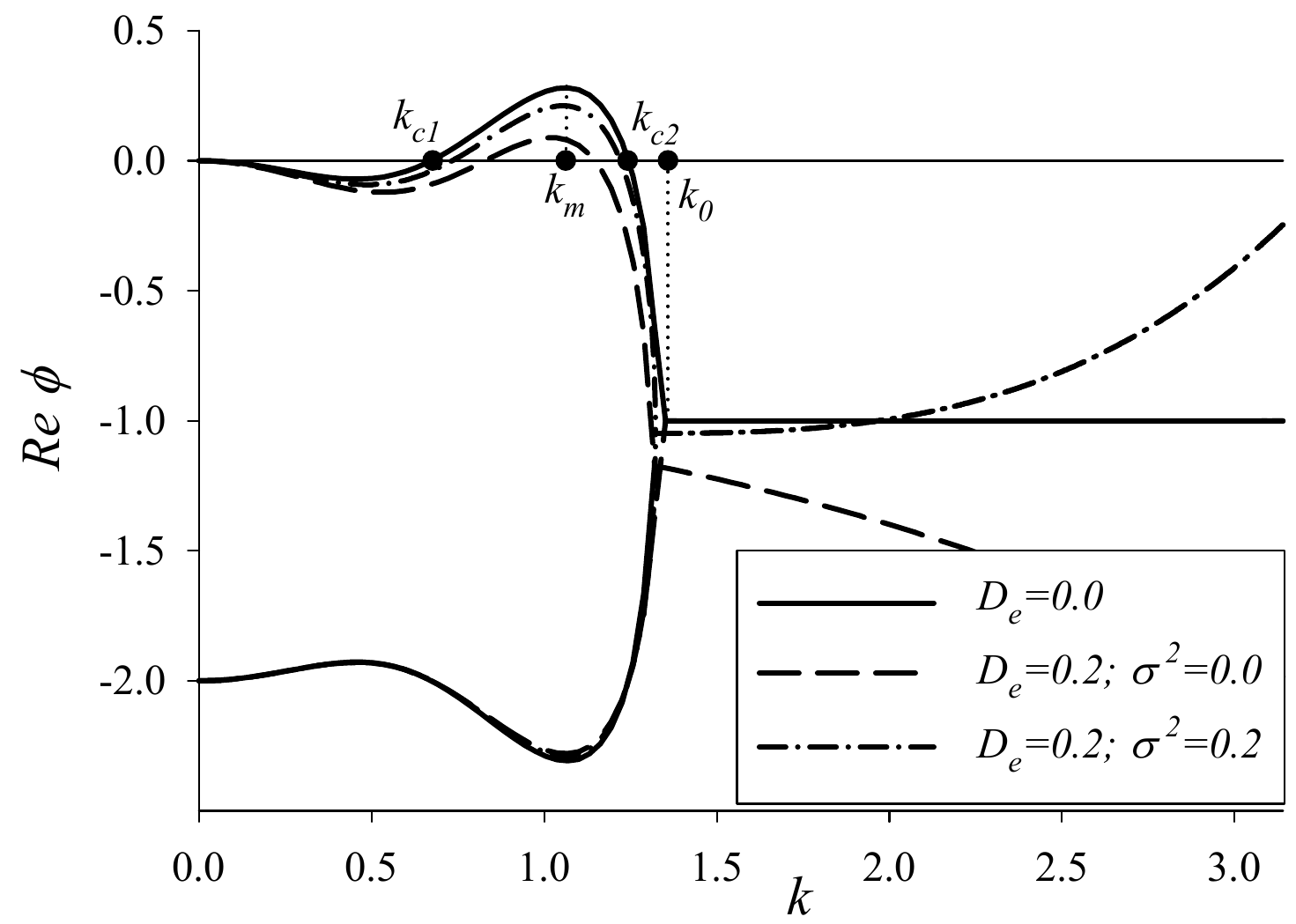} \  b)\includegraphics[width=60mm]{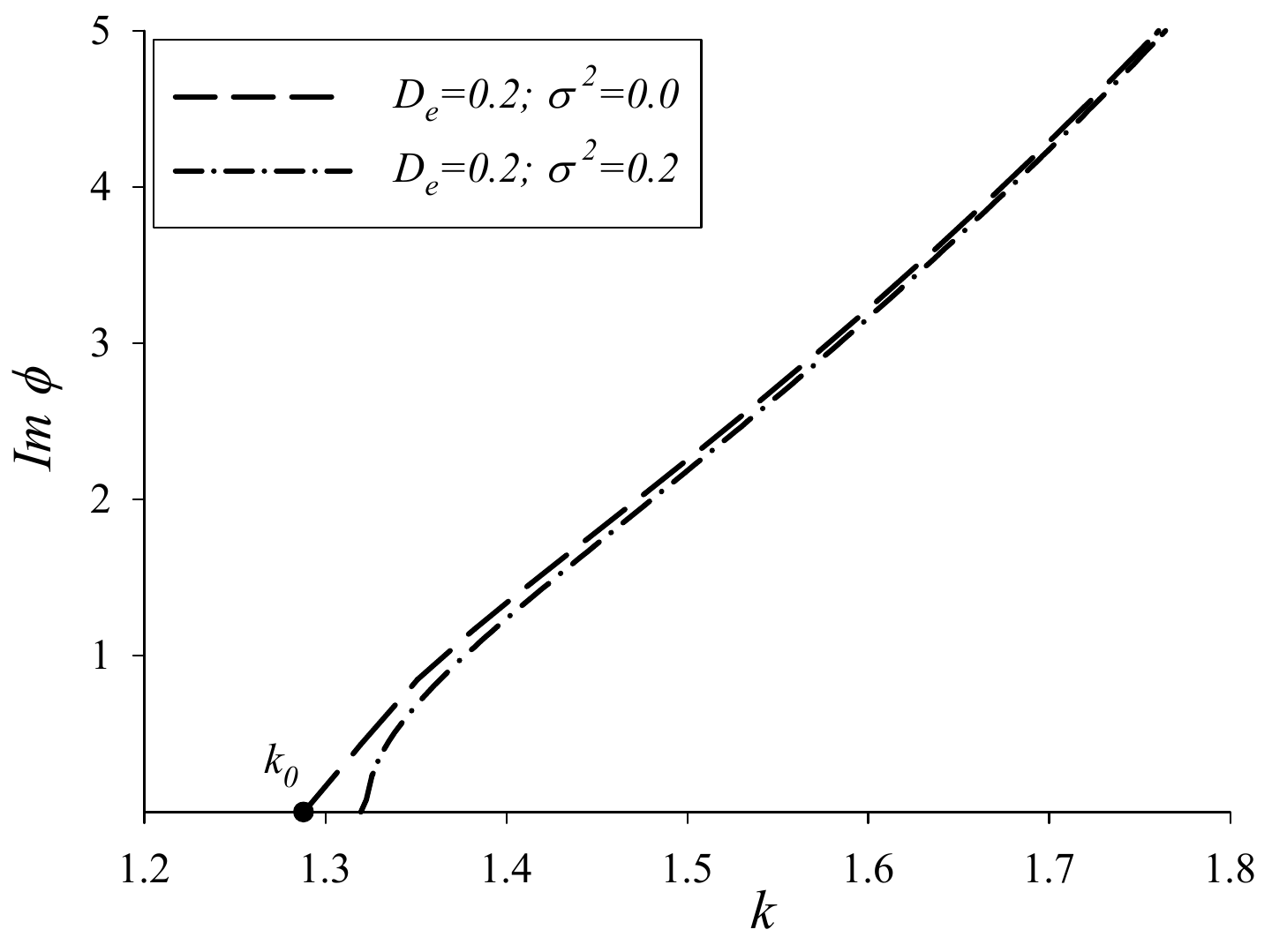}
\caption{Real (a) and imaginary (b) parts of the phase $\phi(k)$ in the
vicinity of the state $x_0=0$ at different values of the external flux
parameters $D_e$ and $\sigma^2$. Other parameters are: $\theta=0.7$,
$\tau_D=0.5$, $r_c=0.5$. \label{ImRePhi0}}
\end{figure}

Let us consider stability of the homogeneous state $x_0=0$ studying real and
imaginary parts of the phase $\phi(k)$ (see Fig.\ref{ImRePhi0}).
From Fig.\ref{ImRePhi0}a it follows that in the simplest case $D_e=0$ (solid
line) unstable modes are limited by the interval of the wave-numbers $k_{c1}\le
k\le k_{c2}$. The deterministic external influence (dashed line) suppresses the
instability progress, resulting to shrinking the domain of unstable modes.
However, external fluctuations act in opposite manner to the deterministic part
of the flux $\mathbf{J}_e$: the domain of unstable modes extends as $\sigma^2$
increases, and oscillating solutions emerge at large $k_0$ (see
Fig.\ref{ImRePhi0}b). Hence, we arrive at competing contributions of regular
and random parts of the external flux. It should be noted that the wave-number
responsible for the most unstable mode and, respectively, for the period of the
formed structure depends on the temperature $\theta$ and the intensity $D_e$.

\begin{figure}[!t]
\centering
 a)\includegraphics[width=50mm]{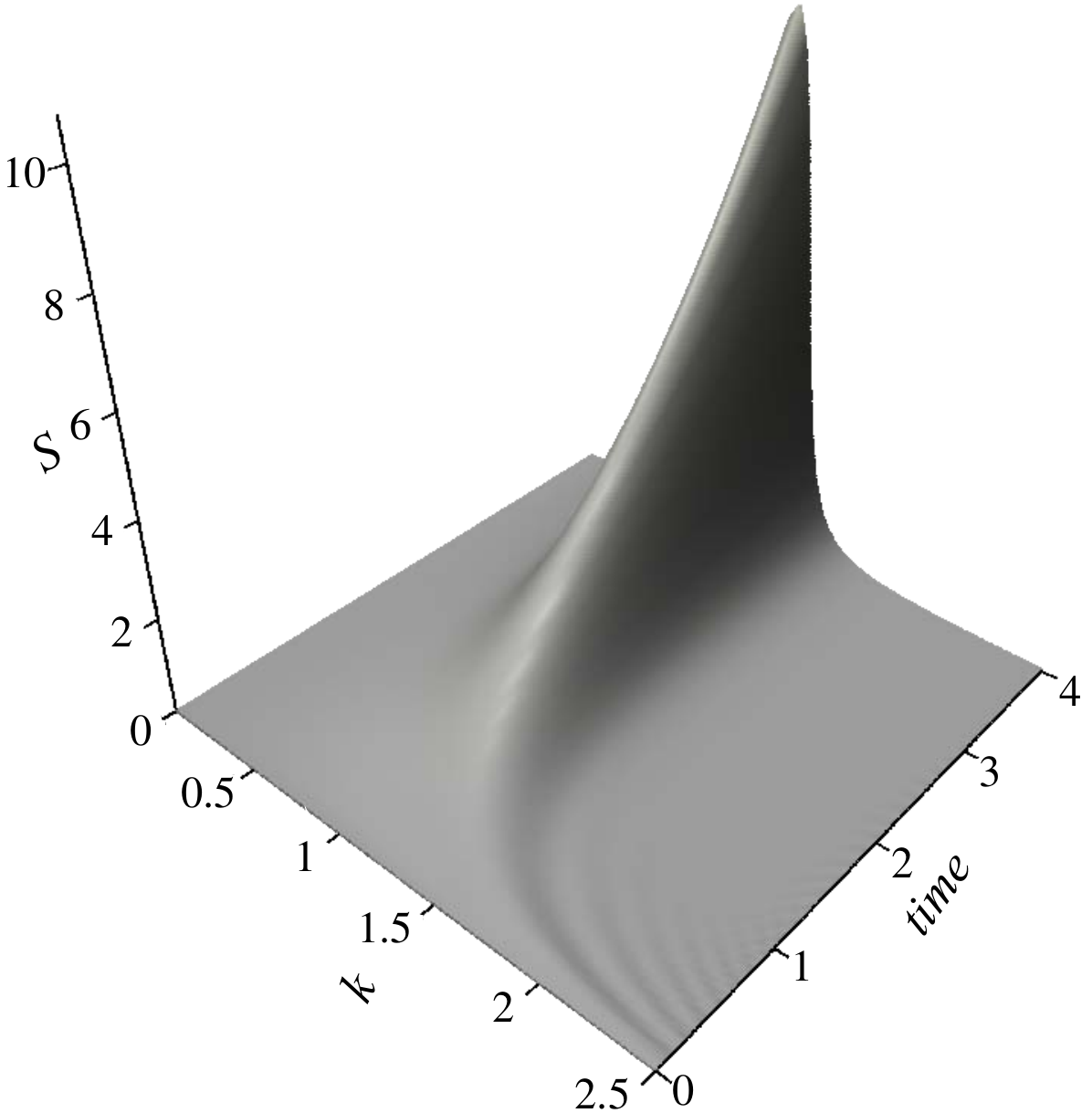}\  b)\includegraphics[width=60mm]{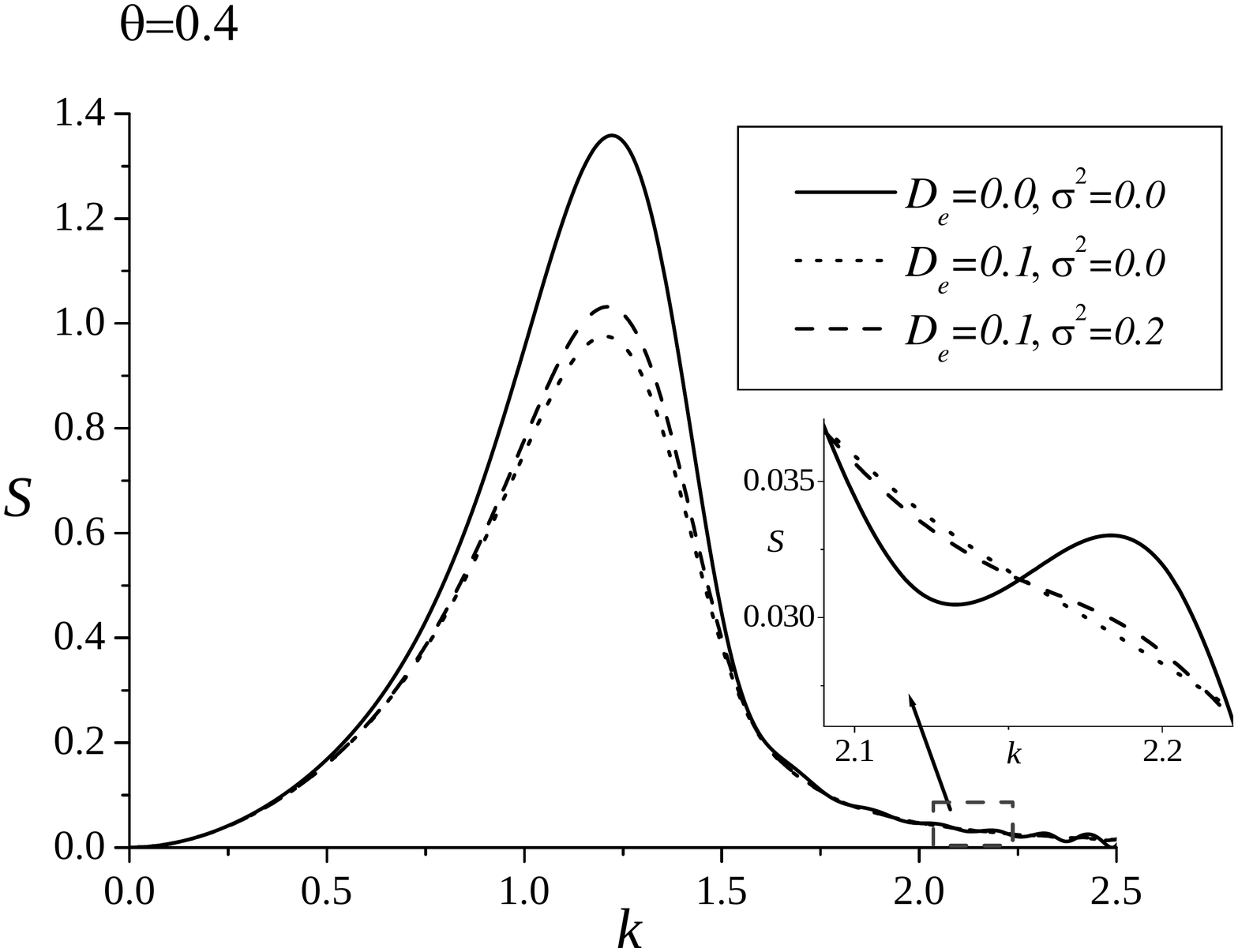}
\caption{Evolution of the structure function at small times at $D_e=0.1$,
$\sigma^2=0.2$ (a) and the structure function at $t=2$ for $\tau_D=1.0$,
$r_c=1.0$, $\theta=0.7$. \label{Skt_es01}}
\end{figure}

Let us study pattern selection processes considering dynamics of the structure
function shown in Fig.\ref{Skt_es01}. It is seen that oscillations in time are
observed at different values of the wave-number $k$. Such a behaviour is
absolutely predictable according to the form of the dynamical equation for the
structure function. More interesting is the wave behavior of $S$ \emph{versus}
the wave-number. The main peak in the dependence $S(k)$ is responsible for the
main period of the structures, whereas other peaks reflects selection of
patterns with small periods. These peaks decay with time growth that means
selection of the one unstable mode $k\simeq k_m$ that gives the main
contribution in the pattern formation processes. The same oscillations can be
found in solutions of the equation for the average $\langle x\rangle$. From
Fig.\ref{Skt_es01}a it is seen that during the system evolution the position of
the main peak is shifted toward the most unstable mode $k\simeq k_m$; a width
of the peak is reduced and it becomes higher. It means that the spatial
patterns become well-defined with sharp interfaces.
From the dependencies $S(k)$ at a given $t$ it follows that the external flux
has a crucial influence on the patterns selection processes. Here the regular
part of the athermal flux ($D_e\ne0$, $\sigma^2=0$) suppresses the pattern
selecting, whereas its stochastic contribution ($\sigma^2\ne0$) results in the
amplification of the magnitude of the structure function peaks sustaining the
pattern selection processes. Let us note that the competition of both regular
and stochastic contributions of the external flux leads to the fact that the
main peak in $S(k)$ dependence decreases at large $D_e$ and its width
increases. It means that the corresponding patterns have more diffuse
interfaces if additional (athermal) diffusion is introduced. However, the
stochastic source of the flux $\mathbf{J}_e$ acts in the opposite manner. The
main peak of the structure function is shifted toward large $k$ as the
temperature $\theta$ decreases.

\begin{figure}[!t]
\centering
 \includegraphics[width=80mm]{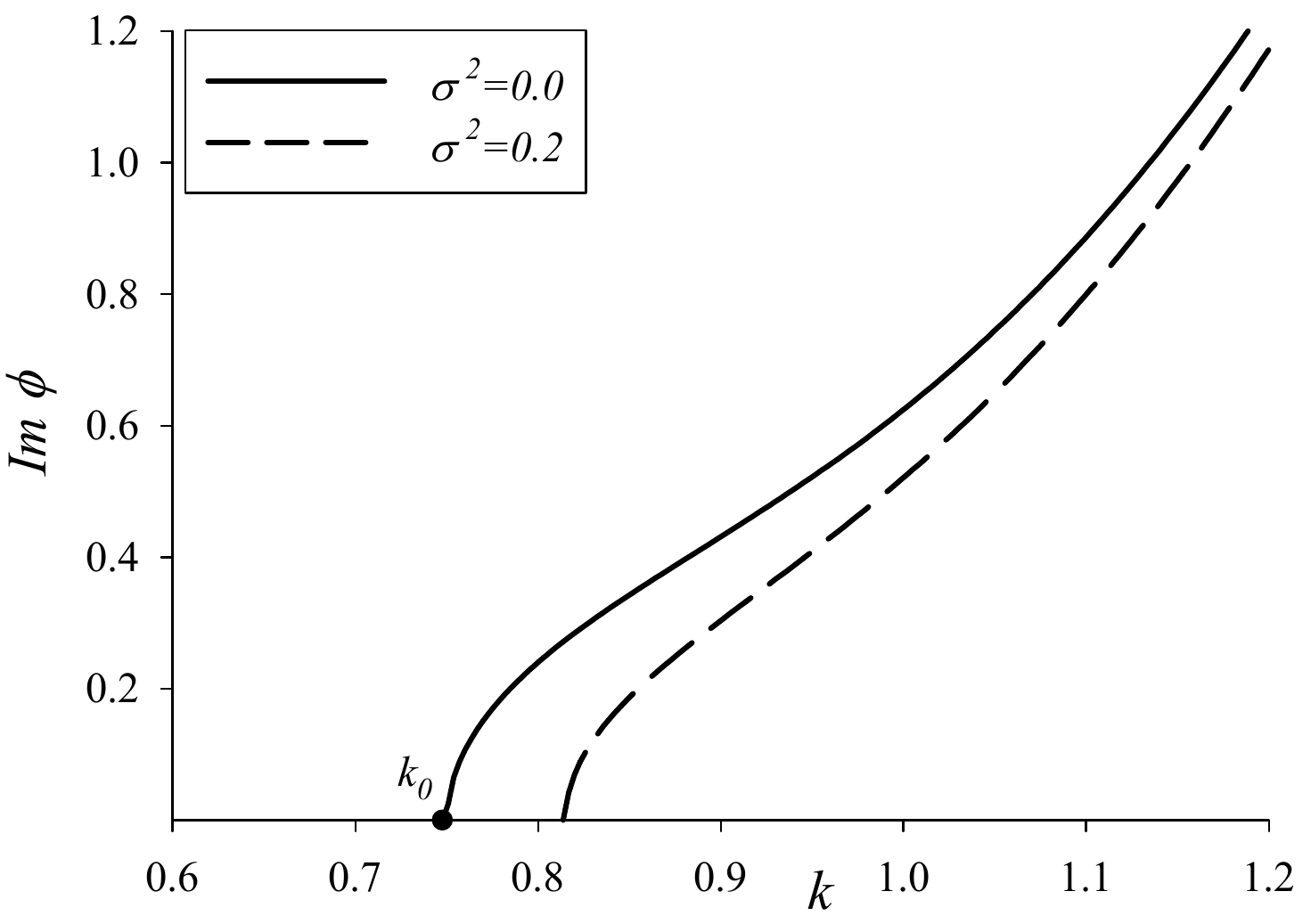}
\caption{Imaginary part of the phase $\phi(k)$ of state $x_0=\sqrt{1-\theta}$
at $\tau_D=0.5$, $r_c=0.5$, $D_e=0.2$, $\theta=0.4$. \label{ImPhinu01}}
\end{figure}

Let us consider stability of the state $x_0=\sqrt{1-\theta}$ at $\theta<1$. We
are interesting in pattern selection processes in the vicinity of the minimum
of the free energy density $f(x)$. From naive considerations one can say that
the system should be stable in the vicinity of the free energy density minimum.
Indeed, here the real part of the phase is negative, $\Re\phi(k)_\pm<0$ always
(not presented here). The imaginary part can exist at $k>k_0$ (see
Fig.\ref{ImPhinu01}). Therefore, in the vicinity of $x_0=\sqrt{1-\theta}$
pattern selection is possible too. Here an increase in the noise intensity
$\sigma^2$ results in a growth of the quantity $k_0$.
\begin{figure}[!t]
\centering
 a)\includegraphics[width=60mm]{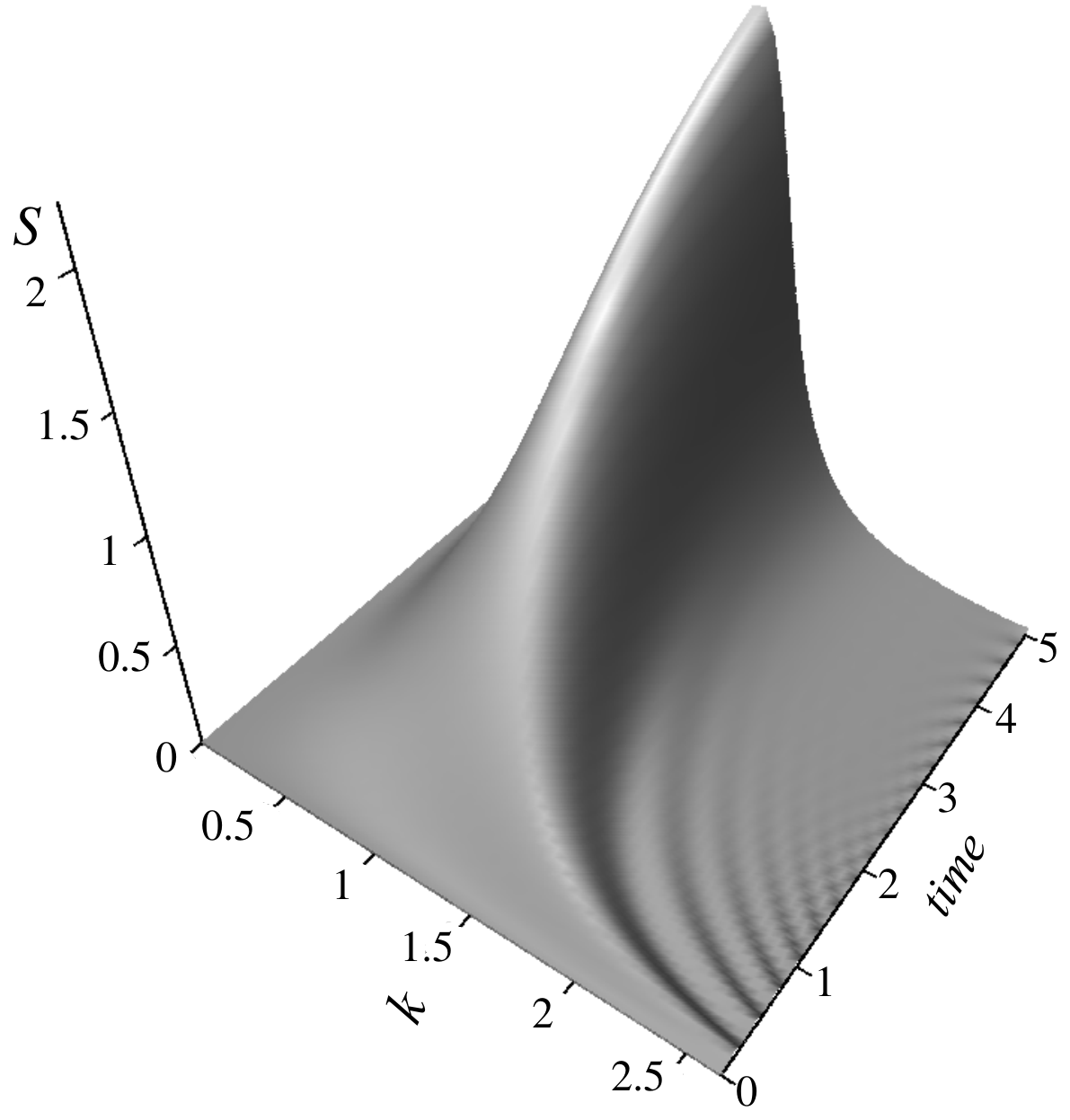} b)\includegraphics[width=60mm]{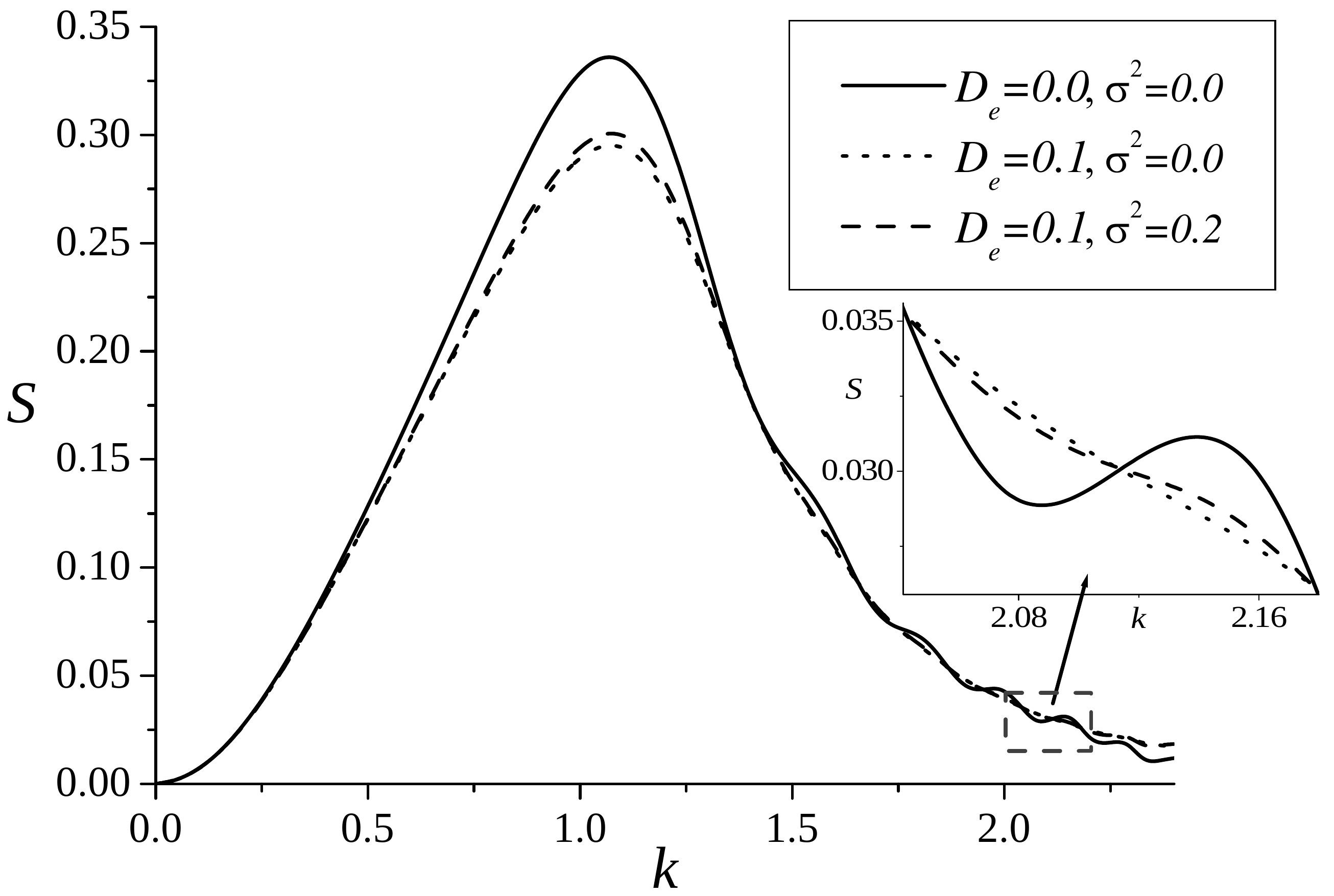}
 \caption{Evolution of the structure function (a) in the vicinity of the state $x_0=\sqrt{1-\theta}$
 at $\sigma^2=0.2$, $D_e=0.1$ and the structure function at different $D_e$ and $\sigma^2$ at $t=2$ (b); $\theta=0.4$, $tau_D=1.0$, $r_c=0.65$.)\label{Skt_os0}}
\end{figure}

The structure function in the vicinity of the state $x_0=\sqrt{1-\theta}$ is
shown in Fig.\ref{Skt_os0} where oscillations \emph{versus} time variable and
wave-number is clearly seen. The position of the main peak of the structure
function (see Fig.\ref{Skt_os0}a) does not change in time at large $t$ and its
magnitude does not increase. It means that patterns are formed rapidly with
sharp interfaces. From Fig.\ref{Skt_os0}b one can see that a height of the main
peak is reduced as $D_e$ increases, the wave behaviour disappears. From another
hand stochastic contribution of the flux $\mathbf{J}_e$ promotes an increase in
the amplitude of the structure function oscillations and increase in the main
peak height. Therefore, due to the stochastic external influence the spatial
patterns become well-defined.

\begin{figure}
\centering
 \includegraphics[width=80mm]{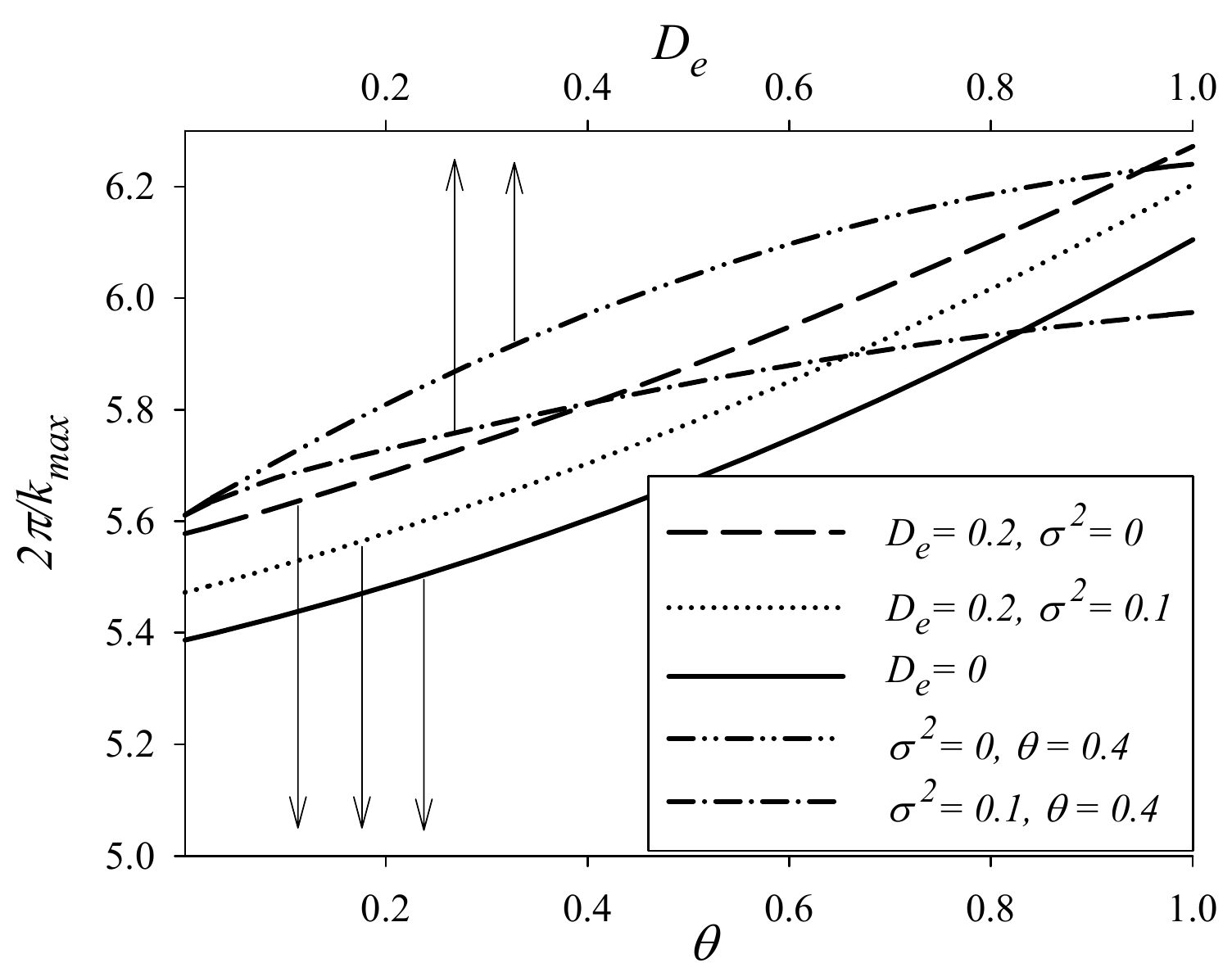}
\caption{Dependencies of the period of patterns $2\pi/k_{max}$ versus
temperature and coefficient $D_e$ and different $\sigma^2$ at $\tau_D=0.5$,
$r_c=0.5$. \label{km}}
\end{figure}
In Fig.\ref{km} we plot the dependence of the period of patterns $2\pi/k_{max}$
computed according to the wave-number $k_{max}$ related to the maximum of the
structure function $S_{max}=S(k_{max})$ at large $t$ when the position of the
main peak does not change. It is seen that the period of structures grows as
temperature and the coefficient $D_e$ increases: thermal heating leads to
formation of large patterns; and additional (athermal) diffusion results to the
same effect due to the constant $D_e$ leads to renormalization of the effective
temperature $\theta_{ef}=1-\theta+D_e$. At large $D_e$ period of patterns at
the fixed $\theta$ increases, but the stochastic external source suppresses
this growth \footnote{Dependencies in Fig.\ref{km} have the same form as
$(k_{m})^{-1}$ \emph{versus} $\theta$ and $D_e$ obtained from the condition of
the maximum of the function $\Re\phi_+(k)$.}.

\section{Simulations}\label{sec3}

\subsection{Discrete representation}

In order to verify the previous predictions, one can perform numerical
simulations of the discrete model (\ref{XJ}) in a square two-dimensional
lattice $N\times N$ of the cell size $\ell$. The partial differential equations
(\ref{XJ}) in the discrete space take the form
\begin{equation}
\begin{split}
 &\frac{{\rm d} x_i}{{\rm d}t}=-(\nabla_R)_{ij}J_{j}+D_e\Delta_{ij}x_j+(\nabla_R)_{ik}\zeta_k(\nabla_L)_{kl}x_l,\\
 \tau_D&\frac{{\rm d} J_i}{{\rm d}t}=-J_i-M(\nabla_L)_{ij}\frac{\partial
 F}{\partial x_j}+\xi_i,\\
 \tau_\zeta&\frac{{\rm d} \zeta_i}{{\rm
 d}t}=-(\delta_{ij}-r_c^2\Delta_{ij})\zeta_j+\tilde\xi_i,
\end{split}
\end{equation}
where index $i$ labels cells, $i=1,\ldots, N^2$; the discrete left and right
operators are introduced as follows
\begin{equation}
\begin{split} &(\nabla_L)_{ij}=\frac{1}{\ell}(\delta_{i,j}-\delta_{i-1,j}),\quad
(\nabla_R)_{ij}=\frac{1}{\ell}(\delta_{i+1,j}-\delta_{i,j}),\\
&(\nabla_L)_{ij}=-(\nabla_R)_{ji},\quad (\nabla_L)_{ij}
(\nabla_R)_{jl}=\Delta_{il}=\frac{1}{\ell^2}(\delta_{i,l+1}-2\delta_{i,l}+\delta_{i,l-1}).
\end{split}
\end{equation}
For the stochastic sources the discrete correlator is of the form
$\langle\xi_i(t)\xi_j(t)\rangle=2\ell^2\sigma^2\delta_{ij}\delta(t-t')$. In the
limit $\tau_\zeta\ll 1$ one has quasi-white noise with
$\langle\tilde\xi_i(t)\rangle=0$,
$\langle\tilde\xi_i(t)\tilde\xi_j(t')\rangle\simeq\delta_{ij}\delta(t-t')$.
Under these conditions we arrive at the stochastic process with properties
given by Eq.(\ref{cz}).

\begin{figure}[!t]
\centering
 \includegraphics[width=100mm]{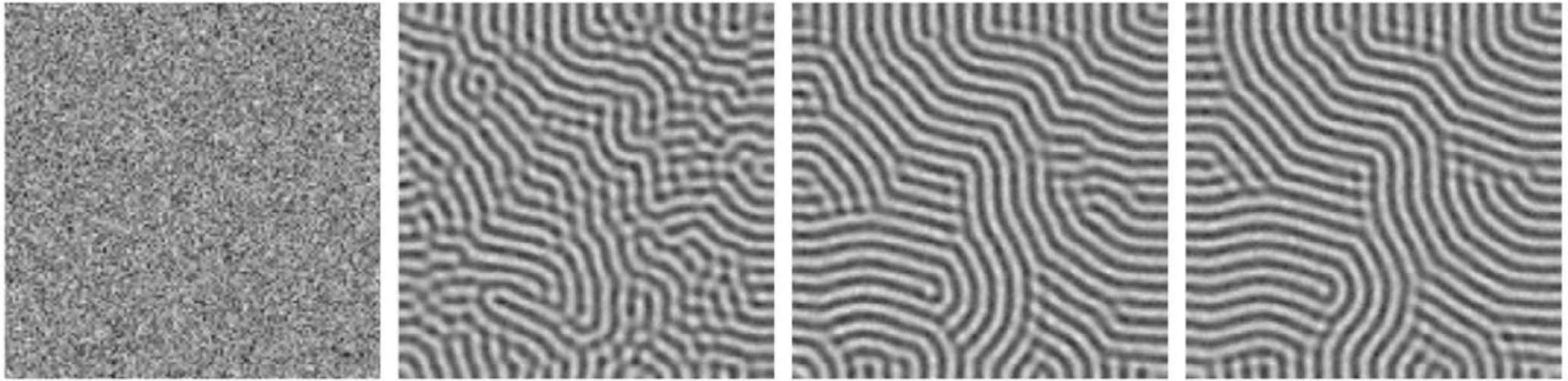}
\caption{Evolution of patterns at $\theta=0.3$, $D_e=0.5$, $\sigma^2=1.0$,
$\tau_D=0.5$, $r_c=1.0$. Snapshots are taken at $t=0$, 50, 500,
1000.\label{evolnu1}}
\end{figure}
Simulations are provided with the time step $\delta t=10^{-3}$ on the square
lattice of the size $128\times 128$ with $\ell=1$. Typical evolution of the
system is shown in Fig.\ref{evolnu1} with initial conditions $\langle
x(\mathbf{r},t=0)\rangle=0$, $\langle(\delta x)^2\rangle=0.3$. It is seen that
linear patterns are formed that corresponds to the standard analysis of the
pattern formation in deterministic systems with such initial conditions
\cite{EKHG_PRL2002}.

\subsection{Analysis of the wave behavior}

To prove the wave behavior of the first statistical moment and the structure
function we analyze the dynamics of two first statistical moments. While the
considered system obeys a mass conservation law ($\int{\rm
d}{\mathbf{r}}x(\mathbf{r},t)=const$, where $const=0$), in our computer
simulations we calculate averages that correspond to positive and negative
values of the stochastic field $x$, i.e. $\langle x \rangle_+=\langle
N^{-2}\sum_i x_i^{>}\rangle$, where $x_i^{>}$ relates to $x_i>0$, and $\langle
x \rangle_-=\langle N^{-2}\sum_i x_i^{<}\rangle$, where $x_i^{<}$ relates to
$x_i<0$, $\langle \ldots\rangle$ means average over experiments. If the
ordering processes realize, then the quantities $\langle x \rangle_+$, $\langle
x \rangle_-$ should increase during the system evolution satisfying the
conservation law $\langle x \rangle\equiv \langle x \rangle_++\langle x
\rangle_-=0$. Another criterion for an existence of an ordered state is growth
of the averaged second moment $J=\langle N^{-2}\sum_i x_i^{2}\rangle$ which in
pattern selection discussion is known as a \emph{convective heat}; it plays a
role of an order parameter in pattern-forming transitions theory \cite{Garcia}.
An alternative definition is $J(t)=\sum_kS(k,t)$, where $S(k,t)$ is a
spherically averaged structure function. Hence $J(t)$ gives the square under
the function $S(k,t)$. Therefore, possible oscillations in $J(t)$ should
manifest the wave behavior of the structure function \emph{versus} time. To
prove the existence of pattern selection processes we analyze the spherically
averaged structure function calculated according to standard definition
$S(k,t)=({N_k})^{-1}\sum_{k\le \mathbf{k}\le k+\Delta k}S_\mathbf{k}(t)$.

\begin{figure}[!t]
\centering
 a) \includegraphics[width=65mm]{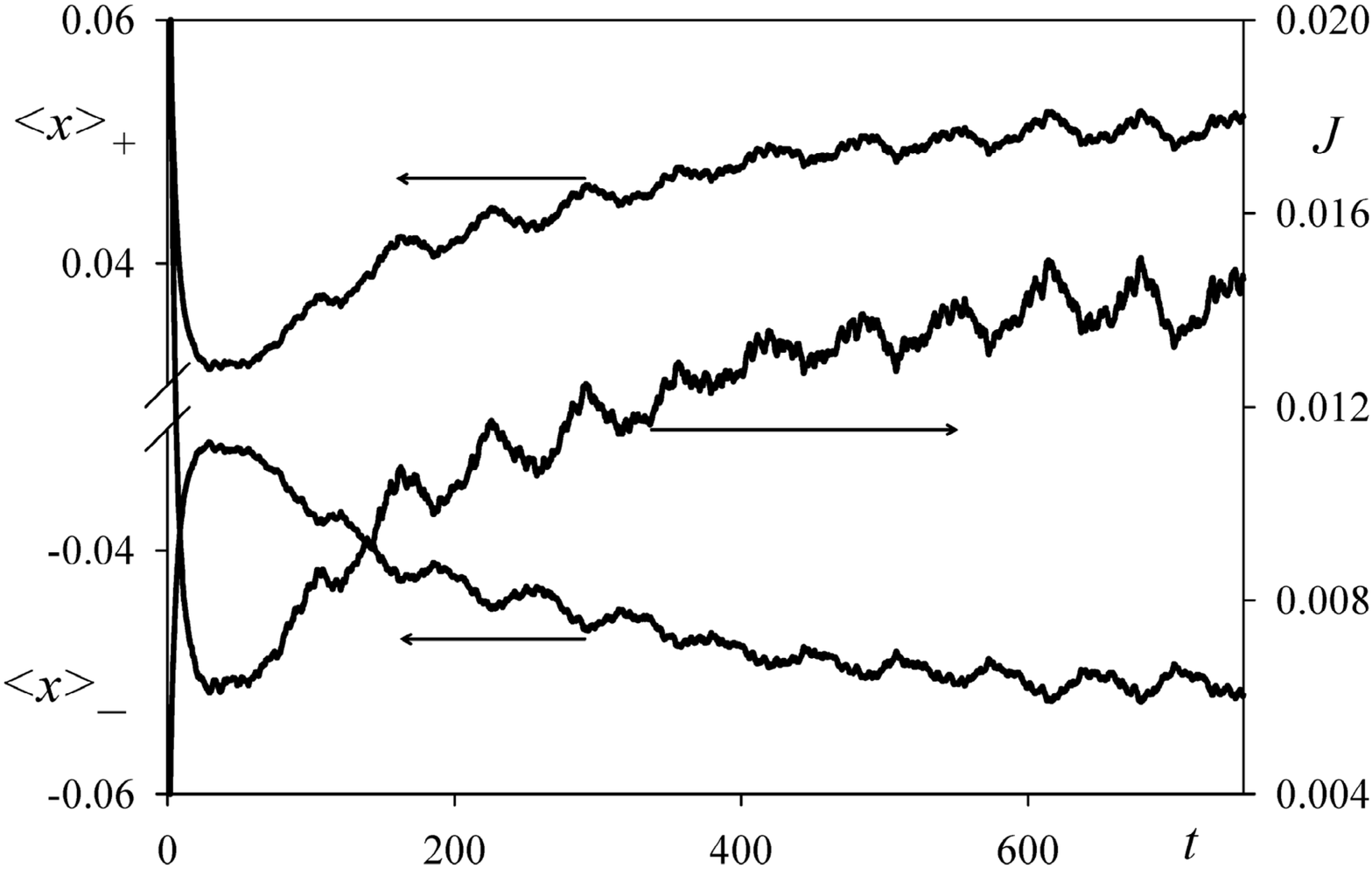} b) \includegraphics[width=65mm]{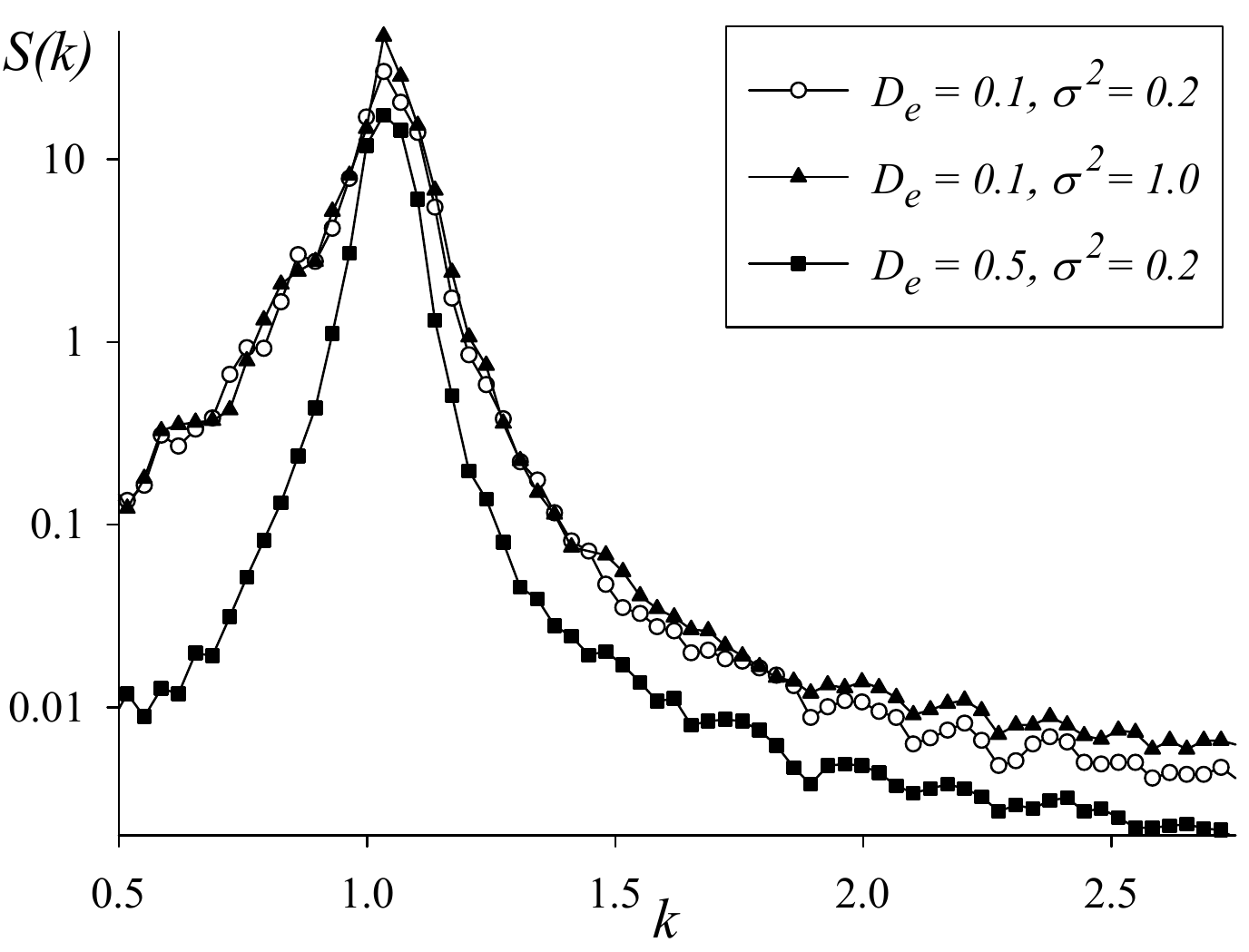}
 \caption{Evolution of averages
$\langle x\rangle_+$, $\langle x\rangle_-$ and the order parameter $J=\langle
x^2 \rangle$ (a) at $\tau_D=0.5$, $r_c=1.0$, $\theta=0.5$, $D_e=0.5$,
$\sigma^2=1.0$ and the spherically averaged structure function $S(k)$ (b) at
$t=1000$, $\tau_D=1.0$, $r_c=1.0$, $\theta=0.2$, $D_e=0.1$,
$\sigma^2=0.2$\label{xpmJ}}
\end{figure}

Results of our computer simulations indicate that the constituents of the total
average $\langle x(t) \rangle_\pm$ and the order parameter $J(t)$ increase
manifesting oscillations (see Fig.\ref{xpmJ}a). Moreover, oscillations of
$\langle x \rangle_+$ and $\langle x \rangle_-$ are realized in an antiphase
manner that satisfies the conservation law. An increase of the order parameter
$J(t)$ means the ordering of the system; the corresponding oscillations
reflects the time oscillations of the structure function. In Fig.\ref{xpmJ}b we
plot the spherically averaged structure function for nonlinear system. It is
seen that the wave behavior \emph{versus} $k$ is realized. With an increase in
the time $t$ additional peaks disappear (not presented here). We plot the
dependence at $\tau_D=1$. At $\tau_D<1$ these peaks are not well pronounced but
exist. Heights of such peaks can be increased setting $\tau_D>1$. In our case
we consider a limit $\tau_D\le 1$ assuming that time scales for the field $x$
and the flux $\mathbf{J}_D$ can be commensurable.

\subsection{Noise-induced pattern-forming transitions}

As was shown above the interactions provided by the Swift-Hohenberg coupling
operator lead to nonequilibrium pattern-forming transitions. In this subsection
we aim to study these transitions by means of standard technique employed in
the equilibrium phase transitions analysis. In the nonequilibrium case the
relative order parameter is the steady state quantity $\eta\equiv
\lim_{t\to\infty}\overline{\langle{J}\rangle}$, where $\overline{\ldots}$ means
average over large time interval. We can use an additional criterion $m_+\equiv
\lim_{t\to\infty}\overline{\langle x(t)\rangle_+}$ to characterize the ordered
state. The extensive related fluctuations are given by the quantity
$\chi=N^{-2}\overline{(\langle J^2\rangle-\langle J\rangle^2)/\langle
J\rangle^2}$ playing a role of the generalized susceptibility. From these
definitions it follows that in the disordered (homogeneous) state one has
$\eta=m_+=0$, in the ordered state the order parameter takes nontrivial values
($\eta\ne0$). In the vicinity of the critical point (for example
$\theta\simeq\tilde\theta_c$) fluctuations grow and the susceptibility $\chi$
increases.

\begin{figure}[!t]
\centering
 \includegraphics[width=65mm]{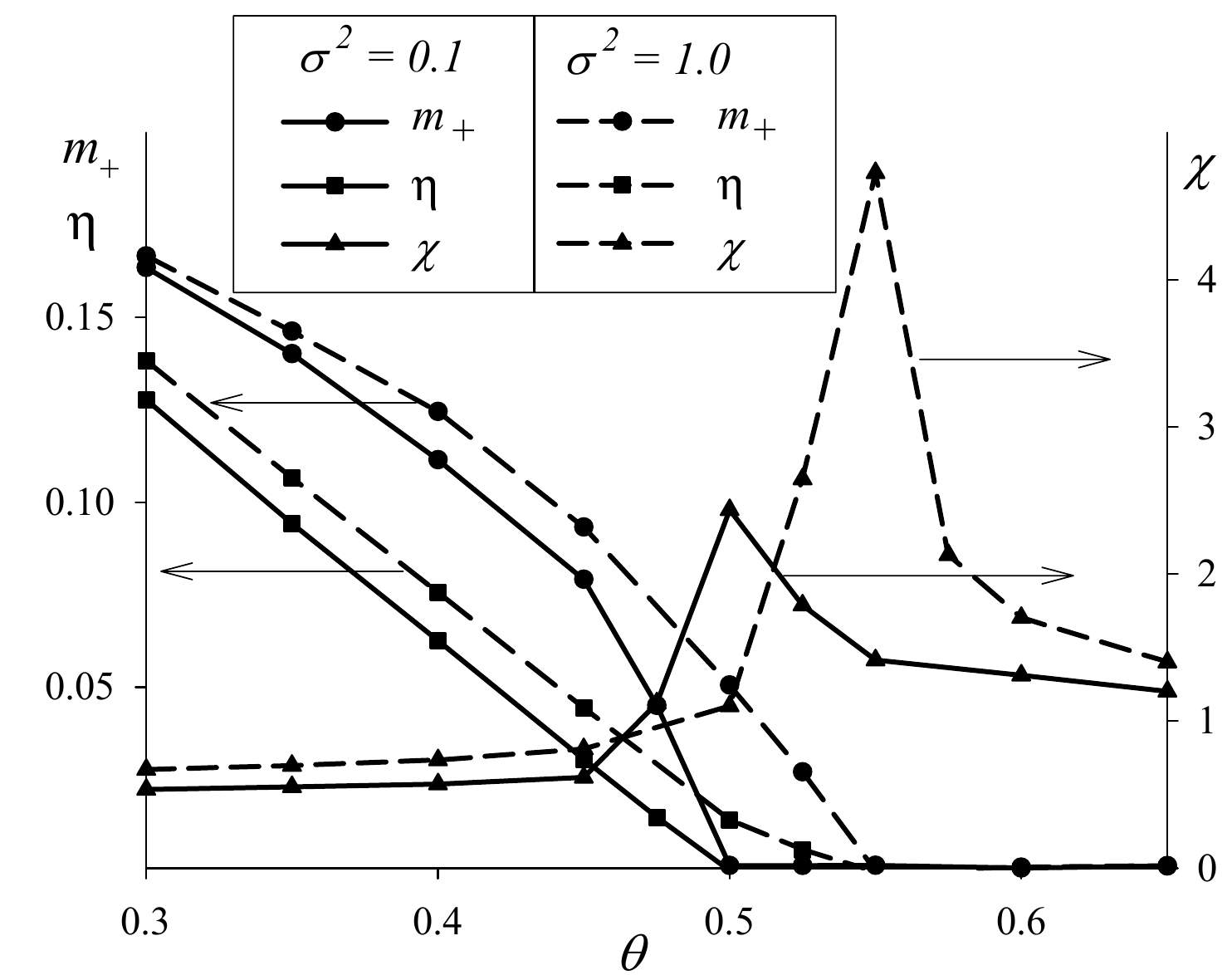} \includegraphics[width=65mm]{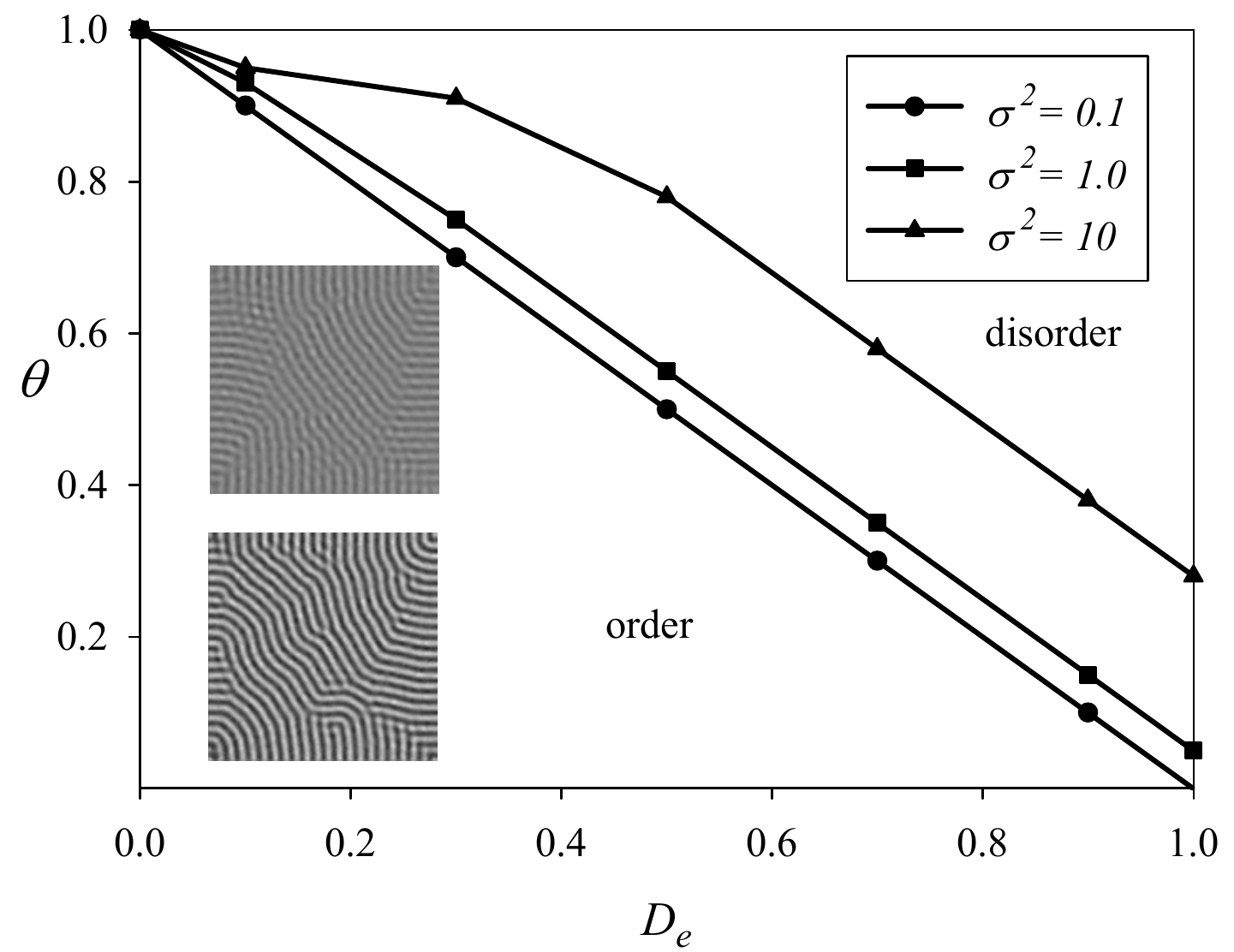}
\caption{Plots of the steady state order parameter $\eta$, the average $m_+$
and the susceptibility $\chi$ (a) at $D_e=0.5$ and phase diagram for
nonequilibrium pattern-forming transitions (b). Other parameters are:
$\tau_D=0.5$, $r_c=1.0$. \label{T(De)}}
\end{figure}

In Fig.\ref{T(De)}а we plot steady state order parameter $\eta$ and the average
$m_+$ \emph{versus} temperature and the parameter $D_e$ at different values of
the external noise intensity $\sigma^2$. It is seen that as the temperature and
ballistic mixing intensity $D_e$ grow the order parameter and the average $m_+$
decreases. In the vicinity of the critical point $\theta\simeq\tilde\theta_c$
the generalized susceptibility shows a well defined maximum at the same
location where the order parameter departs from zero. It clearly seen that
external fluctuations results in renormalization of the critical value for the
control parameter $\varepsilon$. One can find that the critical value for the
noise-induced patterning tends to its mean filed value $\theta_c=1$ as the
noise intensity increases at fixed $D_e$ that decreases $\tilde\theta_c$. To
illustrate the noise-induced shift of the critical point we compute the phase
diagram shown in Fig.\ref{T(De)}b. It follows that at fixed noise intensity
$\sigma^2$ the critical value $\tilde\theta_c$ decreases as the ballistic
mixing intensity $D_e$ grows. One can see that if $\sigma^2$ increases, then
instability of the disordered state emerge at elevated temperatures. This
conclusion is in agreement with the prediction made above from the liner
stability analysis. Moreover, it should be noted that in the domain in the
vicinity of $\tilde\theta_c$ where the fluctuations are large the spatial
patterns have diffuse interfaces, whereas at out off critical temperatures
$\tilde\theta_c$ spatial patterns are well-defined (see insertions in the phase
diagram).

\section{Conclusions}\label{sec5}

We have studied pattern selection processes in periodic stochastic systems with
the hyperbolic transport. Considering the system with thermally sustained flux
and flux of athermal mixing we discuss properties of pattern formation,
selection and nonequilibrium pattern-forming transitions. The dynamics is
studied in terms of both the first statistical moment and the structure
function. Analytical results related to the linear stability analysis are
compared with computer simulations. We have found that external athermal flux
having both regular and stochastic components influences crucially on pattern
selection processes. It was shown that regular part of the external flux
suppresses such processes, whereas it stochastic constituent promotes the
pattern selection. Considering pattern-forming transitions we have shown that
external influence shifts the critical point of the transition where the
regular and random components of the external flux act in competing manner.

\end{document}